\documentclass[aps,twocolumn,pra,tightenlines,floatfix,showpacs]{revtex4}

\usepackage[dvips]{graphicx}
\usepackage[english]{babel}
\usepackage{amsmath}
\usepackage{amssymb}
\usepackage{slashed}

\newcommand{\vect}[1] {\mathbf{#1}}   

\newcommand{\dif} {\mathrm{d}}  

\newcommand{\FT} {\Sigma_2}
\newcommand{\GT} {\Sigma_1}

\newcommand{\bk}{{\mathbf k}}
\newcommand{\w}{\omega}

\newcommand{\Dmat} {G}

\newcommand{\calS}{\mathcal{S}}
\newcommand{\calL}{\mathcal{L}}
\newcommand{\cst}{{\rm const.}}

\newcommand{\fA}{\mathfrak{A}}

\newcommand{\fM}{\mathfrak{M}}
\newcommand{\fS}{\mathfrak{S}}
\newcommand{\fK}{\mathfrak{K}}
\newcommand{\fP}{\mathfrak{P}}
\newcommand{\fL}{\mathfrak{L}}
\allowdisplaybreaks[4]

\begin{document}

\title{ Bosonic thermoelectric transport and breakdown of universality}
\author{A. Ran\c{c}on$^{1}$, Cheng Chin$^{1,2}$, and K. Levin$^{1}$}
\affiliation{$^{1}$James Franck Institute and Department of Physics,
University of Chicago, Chicago, Illinois 60637, USA\\
$^{2}$Enrico Fermi Institute, University of Chicago, Chicago, Illinois 60637, USA}

\date{\today}


\begin{abstract}
We discuss the general principles of transport in normal phase atomic gases,
comparing Bose and Fermi systems. Our study shows that two dimensional bosonic transport
is non-universal with respect to different dissipation mechanisms. Near the superfluid transition temperature
$T_c$, a striking similarity between the fermionic and bosonic transport emerges because super-conducting (fluid)
fluctuation transport for Fermi gases is dominated by the bosonic, Cooper pair component.
As in fluctuation theory, one finds that the Seebeck coefficient changes sign at $T_c$ and
the Lorenz number approaches zero at $T_c$. Our findings appear quantitatively consistent
with recent Bose gas experiments.
\end{abstract}

\maketitle

\section{Introduction to Cold Atom Dissipative Transport}

Cold atom samples differ from electronic systems in
significant ways that offer new opportunities to investigate
transport phenomena \cite{Cheng2,Stadler2012,Brantut2012,Brantut2013,Stoof}. Cold atom systems are highly
versatile: many of them have tunable interactions, can be confined in a single potential well or in
lattices, as well as in different dimensions. Furthermore, at nanoKelvin temperatures, the dynamics
of atoms is very slow (milliseconds), in contrast to the fast dynamics (picoseconds) of electrons in materials. This allows a
detailed scrutiny of atomic motion. One powerful tool in recent cold atom experiments is \emph{in situ} imaging of atoms, which reveals
high space-time resolution images of atomic distributions in snapshots \cite{Gemelke2009,Bakr2009,Sherson2010}. In the condensed matter analogy,
this technique is equivalent to following every electron with a femtosecond temporal resolution.

In further pursuit of understanding both these dynamics and the analogy between
electrons and cold atoms, in this paper 
we apply the theory of dissipative
transport (developed for electrons) to ultracold trapped atoms.
In the last section of the paper we address recent experimental data \cite{Cheng2}.

Dissipative transport 
is to be contrasted with ballistic transport where there are no relevant scattering
processes to limit the transport lifetime. Thus, ballistic transport is observable
when the mean free path is long compared to the relevant dimensions of the system.
We stress that the analogy between electrons and cold atoms
requires careful consideration.
Universal transport laws developed for electrons need to be revisited when applied to bosonic atoms.
In addition, cold atom samples are trapped in a conservative potential and isolated in vacuum. The lack of
thermal and particle reservoirs means total particle and total energy are usually conserved quantities.
A statistical description based on a grand-canonical ensemble may apply only to local observables \cite{landau}.

Furthermore, atoms are neutral and the analogue of electrical current will be the particle
flow or mass flow driven, not by electric field, but by a chemical potential gradient $\nabla \mu$ or temperature gradient $\nabla T$ .
Interestingly, unlike neutral liquid Helium, cold gas superfluids allow the imposition of a non-zero $\nabla \mu$
and thus are rather uniquely amenable to the transport studies we present.

Finally, to employ dissipative, as distinguished from
ballistic transport theory, the samples should be in
the hydrodynamic regime with coherence length much
shorter than the sample and relaxation time less than the
measurement time resolution. When these criteria are satisfied
the atoms can reach a local equilibrium associated with coarse
graining the system over a proper length and time scale. 
Thermodynamic
quantities such as temperature and chemical potential can therefore be defined locally, and their gradients constitute the thermodynamic forces.
This approximation, which we will call the
local equilibrium approximation (LEA), limits the resolution of both temporal and spatial measurements. It should not be confused with
the local density approximation that describes inhomogeneously trapped
atoms in equilibrium. 

Since cold atoms are typically confined in conservative potentials, they are usually free from impurities and background ionic lattices, including their phononic excitations. The fact that atoms are in a clean environment, on the one hand, may simplify many-body calculations; this removes some complexity encountered in electron transport in materials. On the other hand, there
must necessarily be
a relaxation mechanism in order to achieve a steady state and apply
the fundamentals of
transport theory.

In this paper, we discuss the transport behavior of ultracold gases in the normal phase under the assumption of
the LEA, the applicability of which is described in more detail
below. We characterize transport in
two limits where theory is relatively straightforward
and well established. Thus we consider non-interacting
particles as well as (generally interacting) particles
but near superfluid condensation, emphasizing the similarities and differences between Fermi and Bose systems. When the interactions are essentially
negligible, a normal degenerate Fermi gas 
in the dissipative transport regime
has universal thermoelectric properties. By contrast universality 
is absent for the bosonic counterpart. As the system approaches the superfluid critical temperature, the transport tends to be dominated by condensate (superconducting and superfluid) fluctuations in which regime, the Fermi and Bose gases are more similar. It is in this latter regime where there are now (Bose gas) experiments \cite{Cheng2}. Importantly these appear to be reasonably consistent with the theory presented here.

\subsection{Transport Equations }

We define the transport coefficients in terms of the linear response of the particle current $\vect{J}_p$ and the heat current $\vect{J}_Q$ to
temperature gradient $\vect{\nabla} T$ and and chemical potential gradient $\vect\nabla \mu$ as
\begin{eqnarray}
\mathbf{J}_p & =& - L_{11} \nabla \mu ~~- L_{12} \nabla T \\
\mathbf{J}_Q &=& - L_{21} \nabla \mu ~~- L_{22} \nabla T,
\label{eq:1}
\end{eqnarray}

\noindent where $\nabla \mu$ is analogous to the electric field for a charged system. It is convenient
to introduce dimensionless ratios to write this equation in the form (We work with units such that
$\hbar=k_B=1$.)
\begin{equation} \left(\begin{array}{c} \vect{J}_p \\ \vect{J}_Q \\
\end{array}\right)=-\sigma
\left(\begin{array}{cc} 1 & \mathcal{S} \\ \mathcal{P}
& \,\,\,\,\, T\mathcal{L}+\mathcal{S}\mathcal{P} \\ \end{array} \right)
\left(\begin{array}{c} \vect{\nabla}\mu\\ \vect{\nabla}T \\ \end{array}\right),
\label{transportmatrix}
\end{equation}

\noindent where the conductivity  $\sigma \equiv L_{11}$, the Seebeck coefficient (or thermopower) $\mathcal{S}=\mathcal{P}/T\equiv L_{12}/L_{11}$ and the Lorenz
number (which is given by ratio of thermal to mass conductivity, divided
by temperature) $\calL \equiv(L_{22} L_{11}-L_{12} L_{21})/T L_{11}^2$. Note that $L_{12}$ and $L_{21}$ satisfy the
Onsager relation $L_{12}=L_{21}/T$.

\subsection{Kubo versus Boltzmann Approaches}

There are two rather complementary approaches to addressing 
dissipative transport in quantum systems. One can adopt
the approach of linear response theory and apply the Kubo formula, or alternatively one can employ the Boltzmann transport equation which is based on
kinetic theory.  Each has its strengths in accomodating different
physical aspects of the system at hand and the choice for
which to apply is governed by
the underlying goals.
The Kubo formalism is more suitable for imposing conservation principles and sum rules, but it does not generally build in the specifics of the dissipation
mechanism. By contrast, the Boltzmann transport equation focuses on the details of the collision processes leading to dissipation, but is not
as well suited to impose or to verify conservation principles.
For non-interacting gases treated in the Kubo formalism, the results are 
essentially identical to those based on Boltzmann transport equation
at the level of simple relaxation time approximations.

In this paper, we show that Kubo based approaches are well suited to addressing transport of both non-interacting as well as
interacting atomic gases in the near condensation regime. 
Since it is likely
that dissipation in the ultracold gases
is linked to the details of the experimental set up, we will introduce dissipation via a phenomenological parameterization.
The Kubo approach in this form is also suitable for delivering one of the messages of this paper: that in bosonic systems (despite strong evidence for universality in many contexts), transport is highly non-universal and depends on the details of the dissipation.
The philosophy behind our phenomenological approach to dissipation is rather similar to that articulated by Kadanoff and Martin
who in a series of papers emphasized the importance of the Kubo-based correlation functions 
and their symmetries \cite{KadanoffMartin2}. In related work on superfluids \cite{KadanoffMartin}, they argued for the suitability of introducing a phenomenological parameterization of the lifetimes associated with transport.

In this paper we parameterize the relaxation time.
We introduce the quantity
$\tau(\epsilon)$ where $\epsilon$ effectively represents the energy; presumably at
low temperatures $\epsilon\to 0$. 
This is 
treated phenomenologically as $\tau(\epsilon)=\tau_0\epsilon^{\eta/2}$. 
While $\eta$ appears to be arbitrary, in this paper we consider the two different values 
chosen because they have been addressed in the solid state literature:
$\eta=-2$ and $\eta=0$.
The latter is associated with impurity scattering models \cite{Alexandrov}, while
the former is more naturally associated with strictly bosonic
transport as, for example in superconducting fluctuation theories of transport
\cite{VarlamovLarkin,Tinkham}.
Then, reflecting the odd energy or frequency dependene of the boson self energy,
to leading order (in small $\epsilon$) this constrains $\tau(\epsilon)$ and thus $\eta = -2$.

\subsection{Role of Dissipation}
Equilibration is necessary but not sufficient for establishing steady state transport. A relaxation mechanism is needed for equilibration. Such a
mechanism is also central to
many non-equilibrium studies, involving for example,
interaction quenches in the ultracold gases. This was
demonstrated \cite{ourquenchpaper} by our group and in recent experimental 
studies \cite{Hung2011,Makotyn2014}. 
It is, however, important to stress that
mass transport, for example, as in the particle conductivity, cannot exclusively depend on inter-particle collisions as total momentum is conserved in the presence of Galilean invariance. A well behaved transport requires that there be a source of momentum relaxation.

In cold atom experiments dissipation depends on the experimental design, and will arise in a variety of ways. It will be present in thin channel transport
and in experiments with optical lattices or speckle potentials. In addition, there are systematic \cite{Thomas} and intrinsic \cite{Dalibard} fluctuations of the optical potentials that contribute to relaxation. The presence of a harmonic trap alone, which breaks Galilean invariance, is insufficient to avoid an unphysical infinite mass conductivity if there were otherwise no source of relaxation \cite{Wu2014}.

A recent experiment by one of us \cite{Cheng2} has measured the
thermoelectric transport coefficients using three-body recombination loss and heating. These processes lead to locally measurable mass and energy flow or currents. This is in contrast to condensed matter experiments where the temperature and chemical potential (or scalar EM potential) are externally applied. With simple models one can quantify the associated $\nabla \mu$ and $\nabla T$ and thereby deduce transport coefficient ratios. 
This paper aims to address transport more generally
applying to a variety of different experiments, not just to the three-body
recombination loss mechanism. However, it should be stressed that, for this
particular transport scheme, the
LEA will be demonstrated to be appropriate. Most importantly,
along with particle and energy loss, momentum relaxation is present, 
through the loss of particles.

\subsection{Local Equilibrium Approximation}

Indeed, a central assumption of this paper is the applicability of
the local equilibrium approximation.
In dissipative transport theory developed for electrons in solids it
is assumed that in the presence of external fields or perturbations,
a steady state can be achieved. 
This is possible when the atoms 
are in the hydrodynamic regime. Here the coherence length $l_c$ is significantly
less than the system size $l_s$ and thermal relaxation time $\tau$ is much shorter
than the system lifetime $\tau_s$.
One can consider the local
chemical potential gradients $\nabla\mu(x; t)$ and temperature gradients $\nabla T(x; t)$ 
which are essential quantities in transport theory.
That the dynamics of an atomic sample can be
completely described by local thermodynamic variables relies on the rapid
and short-ranged scattering by an external potential, which leads to
relaxation and serves to establish a steady state.
Note that $\nabla\mu(x; t)$ and $\nabla T(x; t)$
can only be defined by coarse graining local variables over a proper length scale $l_m$ and time scale $\tau_m$. This
leads to the constraint $l_c<l_m\ll l_s$ and $\tau<\tau_m\ll\tau_s$ where the subscript $m$ represents the experimental or measurement
variables. When these inequalities hold one is in the regime of
validity of the Local Equilibrium Approximation. This
LEA is a stronger condition than a hydrodynamic approximation since it assumes a history-independent local and short-time thermodynamic equilibrium. 
The LEA, which we assume to be valid throughout this study, will break down for atoms in the collisionless regime or in a smooth trapping potential, where we have $l_c\gg l_s$ and $\tau_m\gg\tau_s$, respectively.

In support of our use of the LEA is the observation that the density profiles are well
described by an equilibrium equation of state
\cite{Gemelke2009}. Also important is the fact that measurements of the coherence
times and correlation lengths (which are discussed in Section V) are consistent
with the constraints, as outlined above.  

\section{Transport for non-interacting bosons and fermions}

The transport properties of weakly interacting, normal Bose and Fermi gases are strikingly
different. In a degenerate Fermi gas, it is well known that the phase space contributions to transport are confined to a narrow energy range around the Fermi energy. By contrast, for Bose gases, there is no such constraint.  As a consequence, the magnitude of transport coefficients $L_{ij}$ for bosons tends to be much larger and  much more sensitive to the detailed assumptions about the nature of the dissipation.

For a Hamiltonian with only one-body terms, we are able to compare the transport properties 
of bosons and fermions using an exact expression  based on the Kubo formula first derived for fermions in Ref.~\onlinecite{Chester1961} but readily generalized to bosons,

\begin{equation}
L_{ij}=T^{1-j}\int_0^\infty {\rm d}\epsilon\, (\epsilon-\mu)^{i+j-2}\frac{2\epsilon}{m d} \rho(\epsilon) \tau(\epsilon) b_\pm^{(1)},
\label{eq:Chester}
\end{equation}

\noindent where $i,j$=1 or 2, $b_\pm^{(1)}\equiv-\partial_{\omega} b_\pm(\omega)$, with $b_\pm(\w)=(z_\pm^{-1}e^{\w/T}\pm 1)^{-1}$ the Fermi/Bose distribution, and the fugacity $z_\pm$ is defined as $z_+=e^{\mu/T}$  for fermions, and $z_-=e^{(\mu-\mu_c)/T}$  for bosons below critical chemical potential $\mu<\mu_c$.
We introduce $\rho(\epsilon)=(m/2\pi)^{d/2}\epsilon^{d/2-1}/\Gamma(d/2)$ the density of states for free particles of mass $m$ in $d$ dimensions (we use units such that $\hbar=k_B=1$). In order to describe a bosonic system at positive chemical potential, one has to include the (non-dissipative) inter-boson interactions, so that
the chemical potential term is given by $\mu-\mu_c$. See for instance 
\cite{Capogrosso-Sansone2010} for an estimate of $\mu_c$ in weak coupling.

The relaxation time introduced earlier 
appears as
$\tau(\epsilon)$ in
Eq.~\eqref{eq:Chester}.
This parameter plays the role of the relaxation time in Boltzmann theories of transport \footnote{In Ref. \onlinecite{Chester1961}, the transport coefficients are proved to be of the form  $L_{ij}=T^{1-j}\int_0^\infty {\rm d}\epsilon\,(\epsilon-\mu)^{i+j-2}G(\epsilon) b_+^{(1)}$, where $G(\epsilon)$ is an unknown function dependent of the details of the Hamiltonian. We can always rewrite $G(\epsilon)=\frac{2\epsilon}{m d} \rho(\epsilon) \tau(\epsilon)$, where $\tau(\epsilon)$ is now unknown. }.

The transport coefficients for non-interacting fermions/bosons can be derived from Eq.~(\ref{eq:Chester}) as
\begin{equation}
\begin{split}
\sigma&= \frac{\Gamma(\zeta+1)\tau(T)}{\Gamma(d/2+1)m\lambda_{\rm dB}^{d}}  |{\rm Li}_{\zeta}(\mp z_\pm)|\\
\calS&= (\zeta+1)\frac{{\rm Li}_{\zeta+1}(\mp z_\pm)}{{\rm Li}_{\zeta}(\mp z_\pm)}-\ln z_\pm\\
\calL&= (\zeta+1)(\zeta+2)\frac{{\rm Li}_{\zeta+2}(\mp z_\pm)}{{\rm Li}_{\zeta}(\mp z_\pm)}-(\calS+\ln z_\pm\big)^2 ,
\label{eq:SSL}
\end{split}
\end{equation}
where $\zeta=(d+\eta)/2$, $\lambda_{\rm dB}=\sqrt{\frac{2\pi}{mT}}$ is the de Brogile length,  and
\begin{equation}
{\rm Li}_{\alpha}(\mp x)=\frac{\mp 1}{\Gamma(\alpha)}\int_0^\infty {\rm d}t \frac{t^{\alpha-1}}{x^{-1}e^{t}\pm 1}
\end{equation}
are the polylogarithm functions.

For degenerate fermions $z_+\gg1$, the factor $b_+^{(1)}$ is strongly peaked at the Fermi energy. Thus the energy dependence of $\tau$ is not relevant and the Lorenz number in this regime is a universal number $\calL_f = \pi^2 /3$ (independent of $\tau$ and $\eta$), known as the 
Wiedermann-Franz law. By contrast, for bosons all energies $\epsilon$ contribute to the integrals, so that $\calL$ depends on the details of the dissipation and has a non-universal behavior.

\begin{figure}[t]
\includegraphics[width=2.6in,clip,angle=-0]
{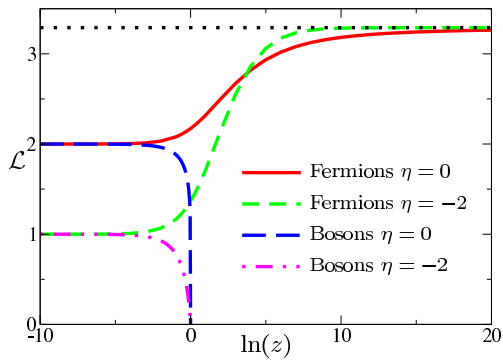}
\caption{
Lorenz number $\calL$ computed from Eq. \eqref{eq:SSL} in $d=2$ with $\eta=-2$ for bosons (solid red lower line) and fermions (short-dashed
green) and with $\eta=0$ for bosons (solid blue upper line) and fermions (dashed magenta line), as function of $\ln(z_\pm)$. The dotted black
line is the Wiedermann-Franz law $\calL=\pi^2/3$ which is reached for fermions in the degeneracy limit $z_+\gg1$. }
\label{fig_coeff}
\end{figure}

Figure \ref{fig_coeff} presents a comparison of the behavior of fermionic and bosonic transport for one particular transport coefficient ratio
(the Lorenz number) $\calL$, for $d=2$ as a function of the log of the fugacity. The figure illustrates how different values of the dissipation
exponents ($\eta=-2$ and $\eta=0$) influence the transport. One sees that for very negative $\ln(z)$, the transport is independent of statistics
and also sensitive to the details of the dissipation. Each component of $\calL$ is larger for the bosonic case, but this effect is
not apparent when plotted as a ratio. One sees that as the fermions cross over to the large positive fugacities, the Lorenz number approaches the universal Wiedermann-Franz law.

While the universality in the fermionic case is evident, this is clearly not the case for bosons. For Bose systems, we observe that even in the quantum
regime the behavior depends on the dissipation mechanism, implying a non-universality of the transport.

\section{Scale invariance of two-dimensional bosonic transport  }

Although, experimentally it has not been
tested in Ref.
\onlinecite{Cheng2}, here we
argue that an scale invariance
should be observed in transport in a two-dimensional Bose gas \footnote{In two-dimensional gas, scale invariance is only approximate due to logarithmic corrections to the scaling.}.
This scaling
arises from physics similar to that found
in the thermodynamics of two-dimensional dilute Bose gases \cite{Hung2011,Yefsah2011}. This scale invariance reflects the
$\mu=T=0$
quantum phase
transition (QPT) between the vacuum and superfluid phase \cite{SachdevBook}.
This zero-temperature phase transition must not be confused with the finite-temperature phase transition between the normal and superfluid phase discussed  in the rest of the paper.
The presence of this QCP implies that a
thermodynamic property such as the pressure is a universal function of the
form
\begin{equation}
P(\mu,T)=\frac{T^{2}}{m}\fP\Big(\frac{\mu}{T},\tilde g(T)\Big).
\end{equation}
Here $\fP(x,y)$ is a universal function (independent of the microscopic details of the system) and $\tilde g(x)$ is a renormalized interaction depending logarithmically on $x$, which
for weakly-interacting bosons can be taken as constant, which we assume in the following. (If
$\tilde g(T)$ varies significantly, the scale invariance is then lost.)  
Under these circumstances
there is scale invariance, so that all thermodynamic
functions in the dilute two-dimensional Bose gas depend only on $\mu/T$ once
the interaction strength has been fixed. The scaling of the pressure is valid in the critical regime close to the QCP, that is $\mu m a^2, Tma^2\ll 1$ \cite{Rancon2012a} ($a$ is the s-wave scattering length). For dilute gases,
this corresponds to the whole $T$-$\mu$ plane in the range of $\mu$ and $T$
relevant for the experiments.

From these results at equilibrium, we can infer that due to this same QCP,
the transport coefficients  also obey scaling relations. Following Ref. \cite{Fisher1990}, one shows that close to a critical point, the conductivity scale as
\begin{equation}
\sigma(\mu,T,g)=s^{2-d}\sigma\Big(\mu s^{1/\nu},T s^{z},\tilde g(s)\Big),
\end{equation}
when lengths are rescaled by a factor $1/s$, where the correlation length exponent $\nu$ and the dynamical  exponent $z$ are given by $1/\nu=z=2$ at the QCP. (See Ref.~\onlinecite{Rancon2012a} for
a detailed discussion of the universal thermodynamics of Bose gases and its relation to the QCP.) 
 Choosing $s=T^{-1/z}$, one obtains
\begin{equation}
\sigma(\mu,T)=T^{(d-2)/z}\fS_d\Big(\frac{\mu}{T},\tilde g(T)\Big),
\end{equation}
where $\fS_d(x,y)=\sigma(x,1,y)$ is a universal function.
This reasoning is easily generalized to the other transport coefficients
\begin{equation}
\begin{split}
L_{12}(\mu,T)&=T^{(d-2)/z}\fA_d\Big(\frac{\mu}{T},\tilde g(T)\Big),\\
L_{22}(\mu,T)&=T^{(d+z-2)/z}\fK_d\Big(\frac{\mu}{T},\tilde g(T)\Big),
\end{split}
\end{equation}
and explicitly in dimension two
\begin{equation}
L_{ij}(\mu,T)=T^{i-1}\calL_{ij}\Big(\frac{\mu}{T},\tilde g\Big),
\end{equation}
 with $\fA_d$, $\fK_d$ and $\calL_{ij}$ universal functions. Note that the scaling arguments are in principle valid only for the singular part of the transport coefficient, and one could expect the presence of a regular part, which would not scale accordingly. However, since all four transport coefficients vanish for all $\mu\leq 0$ at $T=0$ (the system is empty), we can infer that the regular parts are identically zero in the vicinity of the QCP (i.e. for $\mu m a^2, Tma^2\ll 1$).

The normal-superfluid phase transition at finite $T$
(as distinguished from the above vacuum-superfluid zero-temperature phase transition) is
characterized by a singularity of the pressure, which translates into a
singularity of $\fP(x,y)$ at a given $x_c(y)$, where $\fP(x,y)$ is
universal. One
finds that the critical chemical potential (at a
given temperature) is given by $\mu_c(T)=T \fM\big(\tilde g\big)$,
where $\fM(x)$ is also universal.
Building on the universality of $\mu_c(T)/T$, we arrive at
\begin{equation}
L_{ij}(\mu,T)=T^{i-1}\fL_{ij}\Big(\frac{\mu-\mu_c}{T},\tilde g,\eta\Big).
\label{eq_scal_trans}
\end{equation}
Here, we have shown explicitly the dependence of the transport coefficients on the dissipation mechanism through the dissipative coefficient $\eta$  introduced earlier. In particular, this scaling holds far from the normal-superfluid critical regime \footnote{In the critical regime of the normal-superfluid transition, the transport coefficients become scaling functions of $\delta\tilde\mu$ which depend on the finite-temperature fixed point ($O(2)$ Wilson-Fisher in $d=3$ and Berezinskii-Kosterlitz-Thouless in $d=2$), while still respecting the scaling forms given in Eq.  \eqref{eq_scal_trans}.}.

The dependence of the scaling functions on $\eta$, in turn, implies that even though for a \emph{given kind} of dissipation (defined by $\eta$) the transport coefficients of a dilute Bose gas are universal  (\emph{i.e.} described by a function $\fL_{ij}$) in the whole $\mu$-$T$ plane, it will be different for different kinds of dissipation mechanisms (\emph{i.e.} different power laws), thus defining different universality classes \footnote{Note that we work close to equilibrium (the transport coefficient are compute from linear response theory), and thus the scaling functions $\fL_{ij}$ can  be computed at equilibrium for given dissipation mechanisms. }.
It follows directly from Eq.~\eqref{eq_scal_trans} that the Seebeck coefficient (or thermopower) $\calS$ and the
Lorenz ratio $\calL$ are also scale invariant.

\section{Fluctuation transport in Bose and Fermi systems}
\subsection{Fluctuation theory}

In Section II we discussed transport in non-interacting normal gases.
In this context we have seen that there is a dissipative constant parameterized by $\eta$
which is generally unconstrained. Moreover $\eta$ played a central role in determining the transport coefficients of Bose gases,
leading to highly non-universal behavior.
We turn now to Bose and Fermi systems which are normal but near condensation (Bose-Einstein condensation for bosons and the superconducting transition for fermions). This focus stems from recent
Bose gas experiments \cite{Cheng2}.
For the Fermi case, interactions are essential
and one cannot apply the theory of Section II. For the Bose case, the near-condensation regime provides
constraints on the character of dissipation. In this regime we can appeal to theories of fluctuation dynamics
(such as time dependent Landau Ginsburg theory (TDGL)) to constrain the
time dependences and hence frequency dependences of the dissipation. This, in turn, serves to constrain $\eta$,
as
these dynamical theories contain linear time derivatives leading to lifetimes which vary linearly with
frequency or energy.
Relating to the discussion in Section II, interestingly
one arrives at the same
limiting behavior of the Bose transport
coefficients using either
the fluctuation scheme
or Eq.~(\ref{eq:Chester}) with the equivalent $\eta=-2$.

Our focus on normal state transport in the ultracold gases near their condensation temperatures
stems both from recent Bose gas experiments \cite{Cheng2} and also
from the observation that Bose and Fermi superfluids have rather similar properties
in this regime. Both are dominated by a bosonic condensate fluctuation
contribution. In this context \textit{bosonic} transport contributions in superconductors (in the
narrow critical regime) have been
of great interest to the condensed matter community over many decades \cite{VarlamovLarkin,Tinkham}.
Here
one attributes the transport coefficients to fermionic pairs (``composite" bosons) in low
but non-zero momentum states.
These pre-formed pairs yield a greatly enhanced transport in the normal
state, as compared with the behavior deduced from weakly interacting fermion theories such as those in
the previous section.
More recently
there has been a focus on the normal state of high temperature superconductors where it appears
that condensate fluctuation transport may set in at much higher temperatures $T^*$
well above the transition at $T_c$. This reflects the so-called ``pseudogap"
physics \cite{ourreview} and this thought to be related to pre-formed pairs. We  \cite{JS2,ourreview}
and others have argued that this pseudogap should be relevant to Fermi gases at unitarity,
where pairing is strong, but the system is still fermionic.
Thus, if indeed, there is a pseudogap at unitarity, the behavior of transport in
the Fermi gases may serve to reveal it in future experiments.

Transport properties associated with condensate fluctuations are based on two
assumptions. (i) The fluctuations (often called ``Gaussian fluctuations")
represent relatively independent bosons.
At most
one introduces Hartree correlations,  and (ii) 
The dynamics of the
fluctuations incorporates linear time derivatives. Higher order
derivatives are less important due to critical slowing down
effects. This,
importantly, has implications for the frequency dependence of the dissipation
which ultimately affects the behavior of transport.

With these assumptions, the
transport calculations are straightforward and one can evaluate
\cite{ourquenchpaper,Tan2004}
the various transport coefficients following Gaussian fluctuation models
\cite{VarlamovLarkin,Dorsey2,Huse}.
Entering are two factors of the bosonic spectral function, called $A\equiv A(\bk,\w)$:
\begin{align}
\hspace{-1mm}L_{ij}&=\frac{T^{1-j}}{2m^2}\int
\frac{\dif^dk}{(2\pi)^d}\frac{\dif\omega}{2\pi}
\w^{i+j-2} \frac{k^2}{d}\big[A(\mathbf{k},\omega)\big]^2 b_-^{(1)}(\omega).
\label{eq:sigma}
\end{align}
\begin{figure}[t]
\includegraphics[width=1.6in,clip,angle=-90]
{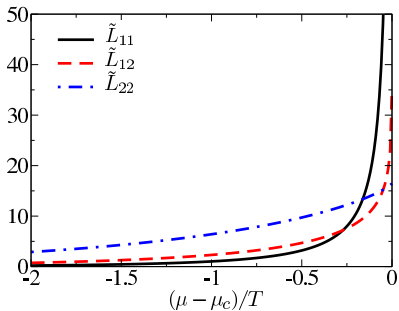}
\caption{Behavior of the transport coefficients $\tilde L_{11}$, $\tilde L_{12}$ and $\tilde L_{22}$ as functions of $(\mu-\mu_c)/T$ for $\Gamma=0.1$ in the Ohmic model ($n=1$) in $d=2$. Their behavior for $(\mu-\mu_c)\to0$ is given by Eq. \eqref{eq_scal}. }
\label{fig2_coeff}
\end{figure}
Here, $b^{(1)}\equiv-\frac{\partial}{\partial\omega} b(\omega)$, with $b(\w)=(e^{\w/T}-1)^{-1}$ the Bose distribution.
We characterize the spectral function in terms of the bosonic propagator
\begin{equation}
\Dmat(\bk,\w) \equiv \Bigl(\omega-\GT(\bk,\omega)-\frac{\bk^2}{2m}+\mu-\mu_c
+\frac{i}{2}\FT(\bk,\omega)\Bigr)^{-1}.
\label{eq:7}
\end{equation}

so that
\begin{equation}\label{eq:A}
A (\bk,\w)=\frac{\FT(\bk,\omega)}{\left(\omega-\frac{\bk^2}{2m}+\mu-\mu_c-\GT(\bk,\omega)\right)^2
+\frac{1}{4}\left(\FT(\bk,\omega)\right)^2},
\end{equation}
where the critical chemical potential $\mu_c$ is a phenomenological parameter, and
we take $\GT(0,0)=0$.
These constraints are convenient, rather than reflective of any deep physics.
We view the effective Hamiltonian (including self energy and dissipation)
as a Hartree-like theory 
\cite{Capogrosso-Sansone2010}
where by definition
$\GT(0,0)=0$ and in this way the superfluid transition occurs at a finite
(positive) chemical potential.

A
crucial feature of bosons is that $\Sigma_2$
changes sign at $\omega = 0$, so that the
spectral function $A$ has the sign of $\omega$.
This is satisfied in
the usual TDGL
dynamics, where at low $\omega$ one has
\begin{equation}
\Sigma_2 \propto \omega^{n}~~~\rm{with}~~n=1,
\end{equation}
corresponding to Ohmic dissipation.
It should be pointed out,
the apparent absence of vertex correction terms
in Eq.~(\ref{eq:sigma}) has been validated in the TDGL literature
(where the bosons only experience Hartree interactions)
and more directly in Ref. \onlinecite{Tan2004} it can be seen
to be consistent with the current conservation constraint
which provides the framework for including vertex terms
in a Kubo formalism.


For the purposes of completeness we briefly revisit the transport nomenclature
found in
superconducting fluctuation
theories,
\cite{Maki,VarlamovLarkin,Tinkham},
where the
fluctuation propagator is found to be proportional to
\begin{equation}
D^{fluc}(\mathbf{q},i \omega) \propto  \frac{1} { [(-1 + i \lambda ) (i \omega) +
\mathcal{D}q^2 +
(8/\pi) (T-T_c)]}
\label{eq:15}
\end{equation}
Note that Eq.~(\ref{eq:15}) can be seen to be equivalent to
Eq.~(\ref{eq:7}), when the dissipation is Ohmic.
Here $\mathcal{D}$ is the so-called diffusion coefficient and
$\lambda = [2 \pi T / g N(0)][N'(0)/N(0)]$
depends on the density of states at the Fermi energy $N(0)$
and its derivative.
By way of further comparison, we note
that in solid state physics one usually describes the approach of the transition by $T-T_c$, whereas $\mu-\mu_c$ is the variable associated with
cold atom experiments, although these two parametrizations are equivalent.

\subsection{Divergences in Transport}

As $\delta\tilde\mu\equiv (\mu -\mu_c)/T$ goes to zero (but away from the
strictly critical regime) we can deduce the transport coefficients
for the two-dimensional case from the integrals in
Eqs.~(\ref{eq:sigma}) and their counterparts.
The literature
is based on the Ohmic case $n=1$ which defines the nature of the divergences
in the $L_{ij}$.
These can be summarized
\cite{Maki,VarlamovLarkin,Tinkham}
in terms of proportionality relations
\begin{equation}
\begin{split}
 \tilde L_{11} &\propto \frac{1}{|\delta\tilde\mu|},\\
\tilde L_{12} &\propto - \ln |\delta\tilde\mu|,\\
\tilde L_{22} &\propto \cst+|\delta\tilde\mu| \ln |\delta\tilde\mu|.
\end{split}
\label{eq_scal}
\end{equation}
From
Eqs. \eqref{eq_scal} we deduce that the Seebeck coefficient (thermopower) and
Lorenz number, which involve ratios of these coefficients, behave as
\begin{equation}
\begin{split}
 \calS &\propto -|\delta\tilde\mu| \ln |\delta\tilde\mu|, \\
\calL &\propto |\delta\tilde\mu|.
\end{split}
\end{equation}
Importantly,
the thermopower changes sign at condensation and the Lorenz number tends to zero linearly.
Recall that $\calL$ must be greater or equal to zero for thermodynamic stability \footnote{In the case $\eta=0$, as used in Ref. \onlinecite{Alexandrov} based on a Boltzmann bosonic approach designed for high-$T_c$ superconductors, one finds $\tilde \sigma\propto \ln|\delta\tilde\mu|$ which is to be contrasted with the Ohmic case  where the divergences at condensation are more evident. As a consequence $\calL \propto  1/\ln |\delta\tilde\mu|$ and the thermopower $\calS$ will similarly vanish logarithmically.}.


In Figure \ref{fig2_coeff} we plot the transport coefficients
for $d=2$ as obtained from the fluctuation theory (Eq. \eqref{eq:sigma})
as a function of scaled chemical potential,
for a typical   $\Gamma=0.1$ \cite{ourquenchpaper}.
We have checked that the value of $\Gamma$ does not change qualitatively the
transport coefficients or power laws for $|\delta\tilde\mu|\lesssim 1$.
This figure shows that the thermal conductivity (related to $L_{22}$) and
the mass conductivity (related to $L_{11}$) as well as the off-diagonal conductivity
satisfying the Onsager relation
($L_{12}=L_{21}/T$) all diverge at the transition. It should be clear, however, that
this divergence is strongest for the mass conductivity.
Important here is that, because of the divergence of $\sigma = L_{11}$,
as $\mu$ goes to $\mu_c$,
$\calL$ (and $\calS$) vanish at $T_c$.

\section{Results: comparison with experiment near condensation  }
We turn now to a comparison of theory and the experiments of Ref.~\onlinecite{Cheng2}. In Ref. \onlinecite{Cheng2}, a Bose gas was confined in an optical trap with peak density $n=5\times10^{13}$cm$^{-3}$, a large scattering length $a=22$~nm, and the temperature  $T= 35$~nK. The trap 
was highly oblate with very strong confinement in the $z-$direction and weak confinement in the two radial directions. The sample is thus in the  quasi-2D regime. For these parameters we are able to address the validity of the LEA. In the normal gas regime, the scattering rate determines the thermalization time scale $\tau_c=\alpha/(4\pi n a^2  v) = 2$~ms where $\alpha=2.7$ is a constant \cite{Monroe1993}, and the coherence length scale is given by the mean-free path $l_c=3.5\mu$m. An optical potential corrugation of up to 3~nK leads to a measured relaxation time of 20~ms of collective excitation. Both these length and time scales are small compared to the the sample size $l_s=60\mu$m$
\gg l_c$ and sample lifetime of $\tau_s=1$s$\gg \tau_c$, which suggests that the LEA is valid over a coarse grained spatial scale of $3.5\mu$m and
time scale $20$~ms.

\begin{figure}[!h]
\includegraphics[width=2.1in,clip]
{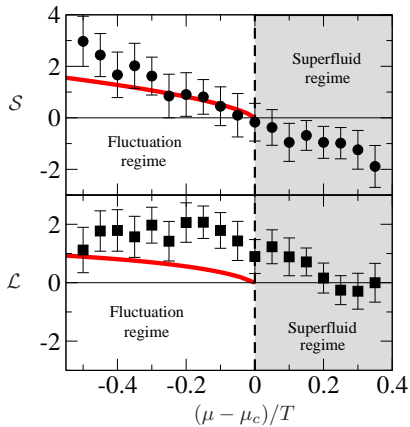}
\caption{Comparison with the experimental data of Ref.\onlinecite{Cheng2} for the thermopower and Lorenz number as functions of $(\mu-\mu_c)/T$ for $\Gamma=0.1$ for the Ohmic model ($n=1$) in $d=2$, see Eq.~\eqref{eq:sigma}.}
\label{fig_L}
\end{figure}

Our comparison between theory and experiment is plotted in
Figs. \ref{fig_L}. Here we show the thermopower (upper panel) and the
Lorenz number (lower panel) for a two-dimensional Bose gas, obtained from Eq.~(\ref{eq:sigma}).
In the simplest terms the data suggests that the thermopower changes sign at
or near the transition. Experimentally, the Lorenz number approaches zero at or
near $T_c$. The error bars are sufficiently large and one
should not infer that $\cal L$ becomes negative on the other side of the transition,
as this is unphysical.
We stress that our theory is applicable only to the normal phase. Indeed, one does
not have many reliable theoretical tools for treating bosons (in any capacity, not
just transport) near, but below the transition.

In the normal phase, the figure shows that the theory and experiment of
Ref.~\onlinecite{Cheng2} coincide rather well for
the thermopower $\calS$, particularly near the condensation
regime. The Lorenz number is off by about
a factor of 2. However, given the simplicity of the model, a
disparity of order unity is not unreasonable.
It should be stressed that the behavior of $\calS$ and $\calL$ are very
strongly reflective of the divergence of the mass conductivity at the phase
transition. The fact that both are found in the data to vanish at or near the transition in the
experiments serves to help validate this experimental methodology 
\cite{Cheng2}.

\section{Conclusions }

Many communities from nuclear physics, astrophysics and condensed matter are interested in
transport in the cold gases. The shear viscosity has been a rather recent focus \cite{Ourviscosity,NJOP,ThomasJLTP08,Qviscosity,
ThomasViscosity} and,
as in this paper, there is an emerging interest  in
thermoelectric properties \cite{Cheng2,Brantut2012}.
There will undoubtedly be many more such experimental transport studies with time.
The theory in this paper is directed towards such experiments. We have emphasized that
the mechanisms for momentum dissipation are very sensitive to the experimental cold atom set up.
This has led us to adopt a Kubo formula approach which treats
dissipation at a non-specific phenomenological level.
An important consequence of our studies
is the observation of quantum critical scaling and
of the non-universality of (bosonic) transport, despite universality
in thermodynamics. This is
in contrast to Fermi liquid or gas systems where
the processes contributing to transport are localized around the Fermi energy, and one finds,
for example a Wiedermann-Franz law for the Lorenz ratio.

Another major finding of this work is the observation that bosonic transport in
the normal state near condensation appears to be semi-quantitatively consistent with observations
and theory in condensed matter studies of (bosonic) superconducting fluctuations 
\cite{VarlamovLarkin,Tinkham}.
In this way
our paper establishes a connection between transport properties in condensed matter and in normal-phase cold atom systems. This connection has other implications for future experiments.
Because
ultracold gases are clean and well controlled, they may help elucidate transport in (fermionic) high $T_c$ systems
\cite{ourreview} where bosonic degrees of freedom (through pre-formed pairs, or vortices or bipolarons, \emph{
etc.}) are believed to dominate transport.
In particular,
high temperature
superconductors in condensed matter
appear to have an anomalously large critical regime where these
fluctuation effects are observed which is associated with the famous ``pseudogap".
In the cold Fermi gases near unitarity there is some debate about whether such a pseudogap exists or not.
This would suggest that future transport studies of fermionic superfluids may help
elucidate this issue.

In contrast to the behavior of degenerate Fermi gases, we reiterate the strong sensitivity of Bose gases to the dissipation mechanism. For bosons, different dissipation mechanisms imply different scaling laws, possibly defining different universality classes. For the future, the universality found in the thermodynamics of dilute Bose gases, either loaded or not in a lattice, \cite{Yefsah2011,Hung2011,Zhang2012,Rancon2012a}, will also need to be addressed in the context of transport. This would allow a direct verification of our prediction of the breakdown of universality in transport.

\vskip1mm
This work is supported by NSF-MRSEC Grant
0820054; CC acknowledges support from NSF PHY-1206095 and ARO-MURI 63834-PH-MUR. We thank
Chih-Chun Chien and Yan He for helpful conversations. CC thanks Grenier Charles and Antoine Georges for useful discussion.

\bibliographystyle{apsrev}
\bibliography{Review,Review4}

\begin{thebibliography}{39}
\expandafter\ifx\csname natexlab\endcsname\relax\def\natexlab#1{#1}\fi
\expandafter\ifx\csname bibnamefont\endcsname\relax
  \def\bibnamefont#1{#1}\fi
\expandafter\ifx\csname bibfnamefont\endcsname\relax
  \def\bibfnamefont#1{#1}\fi
\expandafter\ifx\csname citenamefont\endcsname\relax
  \def\citenamefont#1{#1}\fi
\expandafter\ifx\csname url\endcsname\relax
  \def\url#1{\texttt{#1}}\fi
\expandafter\ifx\csname urlprefix\endcsname\relax\def\urlprefix{URL }\fi
\providecommand{\bibinfo}[2]{#2}
\providecommand{\eprint}[2][]{\url{#2}}

\bibitem[{\citenamefont{Hazlett et~al.}()\citenamefont{Hazlett, Ha, and
  Chin}}]{Cheng2}
\bibinfo{author}{\bibfnamefont{E.~H.} \bibnamefont{Hazlett}},
  \bibinfo{author}{\bibfnamefont{L.-C.} \bibnamefont{Ha}}, \bibnamefont{and}
  \bibinfo{author}{\bibfnamefont{C.}~\bibnamefont{Chin}},
  \bibinfo{note}{eprint, arXiv:1306.4018}.

\bibitem[{\citenamefont{Stadler et~al.}(2012)\citenamefont{Stadler, Krinner,
  Meineke, Brantut, and Esslinger}}]{Stadler2012}
\bibinfo{author}{\bibfnamefont{D.}~\bibnamefont{Stadler}},
  \bibinfo{author}{\bibfnamefont{S.}~\bibnamefont{Krinner}},
  \bibinfo{author}{\bibfnamefont{J.}~\bibnamefont{Meineke}},
  \bibinfo{author}{\bibfnamefont{J.-P.} \bibnamefont{Brantut}},
  \bibnamefont{and}
  \bibinfo{author}{\bibfnamefont{T.}~\bibnamefont{Esslinger}},
  \bibinfo{journal}{Nature} \textbf{\bibinfo{volume}{491}},
  \bibinfo{pages}{736} (\bibinfo{year}{2012}).

\bibitem[{\citenamefont{Brantut et~al.}(2012)\citenamefont{Brantut, Meineke,
  Stadler, Krinner, and Esslinger}}]{Brantut2012}
\bibinfo{author}{\bibfnamefont{J.-P.} \bibnamefont{Brantut}},
  \bibinfo{author}{\bibfnamefont{J.}~\bibnamefont{Meineke}},
  \bibinfo{author}{\bibfnamefont{D.}~\bibnamefont{Stadler}},
  \bibinfo{author}{\bibfnamefont{S.}~\bibnamefont{Krinner}}, \bibnamefont{and}
  \bibinfo{author}{\bibfnamefont{T.}~\bibnamefont{Esslinger}},
  \bibinfo{journal}{Science} \textbf{\bibinfo{volume}{337}},
  \bibinfo{pages}{1069} (\bibinfo{year}{2012}).

\bibitem[{\citenamefont{Brantut et~al.}(2013)\citenamefont{Brantut, Grenier,
  Meineke, Stadler, Krinner, Kollath, Esslinger, and Georges}}]{Brantut2013}
\bibinfo{author}{\bibfnamefont{J.-P.} \bibnamefont{Brantut}},
  \bibinfo{author}{\bibfnamefont{C.}~\bibnamefont{Grenier}},
  \bibinfo{author}{\bibfnamefont{J.}~\bibnamefont{Meineke}},
  \bibinfo{author}{\bibfnamefont{D.}~\bibnamefont{Stadler}},
  \bibinfo{author}{\bibfnamefont{S.}~\bibnamefont{Krinner}},
  \bibinfo{author}{\bibfnamefont{C.}~\bibnamefont{Kollath}},
  \bibinfo{author}{\bibfnamefont{T.}~\bibnamefont{Esslinger}},
  \bibnamefont{and} \bibinfo{author}{\bibfnamefont{A.}~\bibnamefont{Georges}},
  \bibinfo{journal}{Science} \textbf{\bibinfo{volume}{342}},
  \bibinfo{pages}{713} (\bibinfo{year}{2013}).

\bibitem[{\citenamefont{Wong et~al.}(2012)\citenamefont{Wong, van Driel,
  Kittinaradorn, Stoof, and Duine}}]{Stoof}
\bibinfo{author}{\bibfnamefont{C.~H.} \bibnamefont{Wong}},
  \bibinfo{author}{\bibfnamefont{H.~J.} \bibnamefont{van Driel}},
  \bibinfo{author}{\bibfnamefont{R.}~\bibnamefont{Kittinaradorn}},
  \bibinfo{author}{\bibfnamefont{H.~T.~C.} \bibnamefont{Stoof}},
  \bibnamefont{and} \bibinfo{author}{\bibfnamefont{R.~A.} \bibnamefont{Duine}},
  \bibinfo{journal}{Phys. Rev. Lett.} \textbf{\bibinfo{volume}{108}},
  \bibinfo{pages}{075301} (\bibinfo{year}{2012}).

\bibitem[{\citenamefont{Gemelke et~al.}(2009)\citenamefont{Gemelke, Zhang,
  Hung, and Chin}}]{Gemelke2009}
\bibinfo{author}{\bibfnamefont{N.}~\bibnamefont{Gemelke}},
  \bibinfo{author}{\bibfnamefont{X.}~\bibnamefont{Zhang}},
  \bibinfo{author}{\bibfnamefont{C.-L.} \bibnamefont{Hung}}, \bibnamefont{and}
  \bibinfo{author}{\bibfnamefont{C.}~\bibnamefont{Chin}},
  \bibinfo{journal}{Nature} \textbf{\bibinfo{volume}{460}},
  \bibinfo{pages}{995} (\bibinfo{year}{2009}).

\bibitem[{\citenamefont{Bakr et~al.}(2009)\citenamefont{Bakr, Gillen, Peng,
  Folling, and Greiner}}]{Bakr2009}
\bibinfo{author}{\bibfnamefont{W.~S.} \bibnamefont{Bakr}},
  \bibinfo{author}{\bibfnamefont{J.~I.} \bibnamefont{Gillen}},
  \bibinfo{author}{\bibfnamefont{A.}~\bibnamefont{Peng}},
  \bibinfo{author}{\bibfnamefont{S.}~\bibnamefont{Folling}}, \bibnamefont{and}
  \bibinfo{author}{\bibfnamefont{M.}~\bibnamefont{Greiner}},
  \bibinfo{journal}{Nature} \textbf{\bibinfo{volume}{462}}, \bibinfo{pages}{74}
  (\bibinfo{year}{2009}).

\bibitem[{\citenamefont{Sherson et~al.}(2010)\citenamefont{Sherson, Weitenberg,
  Endres, Cheneau, Bloch, and Kuhr}}]{Sherson2010}
\bibinfo{author}{\bibfnamefont{J.~F.} \bibnamefont{Sherson}},
  \bibinfo{author}{\bibfnamefont{C.}~\bibnamefont{Weitenberg}},
  \bibinfo{author}{\bibfnamefont{M.}~\bibnamefont{Endres}},
  \bibinfo{author}{\bibfnamefont{M.}~\bibnamefont{Cheneau}},
  \bibinfo{author}{\bibfnamefont{I.}~\bibnamefont{Bloch}}, \bibnamefont{and}
  \bibinfo{author}{\bibfnamefont{S.}~\bibnamefont{Kuhr}},
  \bibinfo{journal}{Nature} \textbf{\bibinfo{volume}{467}}, \bibinfo{pages}{68}
  (\bibinfo{year}{2010}).

\bibitem[{\citenamefont{Landau and Lifshits}(1980)}]{landau}
\bibinfo{author}{\bibfnamefont{L.~D.} \bibnamefont{Landau}} \bibnamefont{and}
  \bibinfo{author}{\bibfnamefont{E.~M.} \bibnamefont{Lifshits}},
  \emph{\bibinfo{title}{Statistical Physics}}
  (\bibinfo{publisher}{Butterworth-Heinemann}, \bibinfo{address}{Oxford},
  \bibinfo{year}{1980}).

\bibitem[{\citenamefont{Kadanoff and Martin}(1963)}]{KadanoffMartin2}
\bibinfo{author}{\bibfnamefont{L.~P.} \bibnamefont{Kadanoff}} \bibnamefont{and}
  \bibinfo{author}{\bibfnamefont{P.~C.} \bibnamefont{Martin}},
  \bibinfo{journal}{Ann. of Phys.} \textbf{\bibinfo{volume}{24}},
  \bibinfo{pages}{419} (\bibinfo{year}{1963}).

\bibitem[{\citenamefont{Kadanoff and Martin}(1961)}]{KadanoffMartin}
\bibinfo{author}{\bibfnamefont{L.~P.} \bibnamefont{Kadanoff}} \bibnamefont{and}
  \bibinfo{author}{\bibfnamefont{P.~C.} \bibnamefont{Martin}},
  \bibinfo{journal}{Phys. Rev.} \textbf{\bibinfo{volume}{124}},
  \bibinfo{pages}{670} (\bibinfo{year}{1961}).

\bibitem[{\citenamefont{Alexandrov and Mott}(1993)}]{Alexandrov}
\bibinfo{author}{\bibfnamefont{A.~S.} \bibnamefont{Alexandrov}}
  \bibnamefont{and} \bibinfo{author}{\bibfnamefont{N.~F.} \bibnamefont{Mott}},
  \bibinfo{journal}{Phys. Rev. Lett.} \textbf{\bibinfo{volume}{71}},
  \bibinfo{pages}{1075} (\bibinfo{year}{1993}).

\bibitem[{\citenamefont{Larkin and Varlamov}(2005)}]{VarlamovLarkin}
\bibinfo{author}{\bibfnamefont{A.}~\bibnamefont{Larkin}} \bibnamefont{and}
  \bibinfo{author}{\bibfnamefont{A.}~\bibnamefont{Varlamov}},
  \emph{\bibinfo{title}{Theory of Fluctuations in Superconductors}}
  (\bibinfo{publisher}{Oxford University Press}, \bibinfo{year}{2005}).

\bibitem[{\citenamefont{Skocpol and Tinkham}(1975)}]{Tinkham}
\bibinfo{author}{\bibfnamefont{W.~J.} \bibnamefont{Skocpol}} \bibnamefont{and}
  \bibinfo{author}{\bibfnamefont{M.}~\bibnamefont{Tinkham}},
  \bibinfo{journal}{Rep. Prog. Phys.} \textbf{\bibinfo{volume}{38}},
  \bibinfo{pages}{1049} (\bibinfo{year}{1975}).

\bibitem[{\citenamefont{Ran\c{c}on et~al.}(2013)\citenamefont{Ran\c{c}on, Hung,
  Chin, and Levin}}]{ourquenchpaper}
\bibinfo{author}{\bibfnamefont{A.}~\bibnamefont{Ran\c{c}on}},
  \bibinfo{author}{\bibfnamefont{C.-L.} \bibnamefont{Hung}},
  \bibinfo{author}{\bibfnamefont{C.}~\bibnamefont{Chin}}, \bibnamefont{and}
  \bibinfo{author}{\bibfnamefont{K.}~\bibnamefont{Levin}},
  \bibinfo{journal}{Phys. Rev. A} \textbf{\bibinfo{volume}{88}},
  \bibinfo{pages}{031601(R)} (\bibinfo{year}{2013}).

\bibitem[{\citenamefont{Hung et~al.}(2011)\citenamefont{Hung, Zhang, Gemelke,
  and Chin}}]{Hung2011}
\bibinfo{author}{\bibfnamefont{C.-L.} \bibnamefont{Hung}},
  \bibinfo{author}{\bibfnamefont{X.}~\bibnamefont{Zhang}},
  \bibinfo{author}{\bibfnamefont{N.}~\bibnamefont{Gemelke}}, \bibnamefont{and}
  \bibinfo{author}{\bibfnamefont{C.}~\bibnamefont{Chin}},
  \bibinfo{journal}{Nature} \textbf{\bibinfo{volume}{470}},
  \bibinfo{pages}{236} (\bibinfo{year}{2011}).

\bibitem[{\citenamefont{Makotyn et~al.}(2014)\citenamefont{Makotyn, Klauss,
  Goldberger, Cornell, and Jin}}]{Makotyn2014}
\bibinfo{author}{\bibfnamefont{P.}~\bibnamefont{Makotyn}},
  \bibinfo{author}{\bibfnamefont{C.~E.} \bibnamefont{Klauss}},
  \bibinfo{author}{\bibfnamefont{D.~L.} \bibnamefont{Goldberger}},
  \bibinfo{author}{\bibfnamefont{E.~A.} \bibnamefont{Cornell}},
  \bibnamefont{and} \bibinfo{author}{\bibfnamefont{D.~S.} \bibnamefont{Jin}},
  \bibinfo{journal}{Nature Physics} \textbf{\bibinfo{volume}{10}},
  \bibinfo{pages}{116} (\bibinfo{year}{2014}).

\bibitem[{\citenamefont{Savard et~al.}(1997)\citenamefont{Savard, O'Hara, and
  Thomas}}]{Thomas}
\bibinfo{author}{\bibfnamefont{T.~A.} \bibnamefont{Savard}},
  \bibinfo{author}{\bibfnamefont{K.~M.} \bibnamefont{O'Hara}},
  \bibnamefont{and} \bibinfo{author}{\bibfnamefont{J.~E.}
  \bibnamefont{Thomas}}, \bibinfo{journal}{Phys. Rev. A}
  \textbf{\bibinfo{volume}{56}}, \bibinfo{pages}{R1095} (\bibinfo{year}{1997}).

\bibitem[{\citenamefont{Dalibard and Cohen-Tannoudji}(1985)}]{Dalibard}
\bibinfo{author}{\bibfnamefont{J.}~\bibnamefont{Dalibard}} \bibnamefont{and}
  \bibinfo{author}{\bibfnamefont{C.}~\bibnamefont{Cohen-Tannoudji}},
  \bibinfo{journal}{J. Opt. Soc. Am. B} \textbf{\bibinfo{volume}{2}},
  \bibinfo{pages}{1707} (\bibinfo{year}{1985}).

\bibitem[{\citenamefont{Wu and Zaremba}(2014)}]{Wu2014}
\bibinfo{author}{\bibfnamefont{Z.}~\bibnamefont{Wu}} \bibnamefont{and}
  \bibinfo{author}{\bibfnamefont{E.}~\bibnamefont{Zaremba}},
  \bibinfo{journal}{Annals of Physics} \textbf{\bibinfo{volume}{342}},
  \bibinfo{pages}{214 } (\bibinfo{year}{2014}).

\bibitem[{\citenamefont{Chester and Thellung}(1961)}]{Chester1961}
\bibinfo{author}{\bibfnamefont{G.~V.} \bibnamefont{Chester}} \bibnamefont{and}
  \bibinfo{author}{\bibfnamefont{A.}~\bibnamefont{Thellung}},
  \bibinfo{journal}{Proceedings of the Physical Society}
  \textbf{\bibinfo{volume}{77}}, \bibinfo{pages}{1005} (\bibinfo{year}{1961}).

\bibitem[{\citenamefont{Capogrosso-Sansone
  et~al.}(2010)\citenamefont{Capogrosso-Sansone, Giorgini, Pilati, Pollet,
  Prokof'ev, Svistunov, and Troyer}}]{Capogrosso-Sansone2010}
\bibinfo{author}{\bibfnamefont{B.}~\bibnamefont{Capogrosso-Sansone}},
  \bibinfo{author}{\bibfnamefont{S.}~\bibnamefont{Giorgini}},
  \bibinfo{author}{\bibfnamefont{S.}~\bibnamefont{Pilati}},
  \bibinfo{author}{\bibfnamefont{L.}~\bibnamefont{Pollet}},
  \bibinfo{author}{\bibfnamefont{N.}~\bibnamefont{Prokof'ev}},
  \bibinfo{author}{\bibfnamefont{B.}~\bibnamefont{Svistunov}},
  \bibnamefont{and} \bibinfo{author}{\bibfnamefont{M.}~\bibnamefont{Troyer}},
  \bibinfo{journal}{New Journal of Physics} \textbf{\bibinfo{volume}{12}},
  \bibinfo{pages}{043010} (\bibinfo{year}{2010}).

\bibitem[{\citenamefont{Yefsah et~al.}(2011)\citenamefont{Yefsah, Desbuquois,
  Chomaz, G\"unter, and Dalibard}}]{Yefsah2011}
\bibinfo{author}{\bibfnamefont{T.}~\bibnamefont{Yefsah}},
  \bibinfo{author}{\bibfnamefont{R.}~\bibnamefont{Desbuquois}},
  \bibinfo{author}{\bibfnamefont{L.}~\bibnamefont{Chomaz}},
  \bibinfo{author}{\bibfnamefont{K.~J.} \bibnamefont{G\"unter}},
  \bibnamefont{and} \bibinfo{author}{\bibfnamefont{J.}~\bibnamefont{Dalibard}},
  \bibinfo{journal}{Phys. Rev. Lett.} \textbf{\bibinfo{volume}{107}},
  \bibinfo{pages}{130401} (\bibinfo{year}{2011}).

\bibitem[{\citenamefont{Sachdev}(2011)}]{SachdevBook}
\bibinfo{author}{\bibfnamefont{S.}~\bibnamefont{Sachdev}},
  \emph{\bibinfo{title}{Quantum Phase Transitions}}
  (\bibinfo{publisher}{Cambridge University Press},
  \bibinfo{address}{Cambridge, England}, \bibinfo{year}{2011}),
  \bibinfo{edition}{2nd} ed.

\bibitem[{\citenamefont{Ran\c{c}on and Dupuis}(2012)}]{Rancon2012a}
\bibinfo{author}{\bibfnamefont{A.}~\bibnamefont{Ran\c{c}on}} \bibnamefont{and}
  \bibinfo{author}{\bibfnamefont{N.}~\bibnamefont{Dupuis}},
  \bibinfo{journal}{Phys. Rev. A} \textbf{\bibinfo{volume}{85}},
  \bibinfo{pages}{063607} (\bibinfo{year}{2012}).

\bibitem[{\citenamefont{Fisher et~al.}(1990)\citenamefont{Fisher, Grinstein,
  and Girvin}}]{Fisher1990}
\bibinfo{author}{\bibfnamefont{M.~P.~A.} \bibnamefont{Fisher}},
  \bibinfo{author}{\bibfnamefont{G.}~\bibnamefont{Grinstein}},
  \bibnamefont{and} \bibinfo{author}{\bibfnamefont{S.~M.}
  \bibnamefont{Girvin}}, \bibinfo{journal}{Phys. Rev. Lett.}
  \textbf{\bibinfo{volume}{64}}, \bibinfo{pages}{587} (\bibinfo{year}{1990}).

\bibitem[{\citenamefont{Chen et~al.}(2005)\citenamefont{Chen, Stajic, Tan, and
  Levin}}]{ourreview}
\bibinfo{author}{\bibfnamefont{Q.~J.} \bibnamefont{Chen}},
  \bibinfo{author}{\bibfnamefont{J.}~\bibnamefont{Stajic}},
  \bibinfo{author}{\bibfnamefont{S.~N.} \bibnamefont{Tan}}, \bibnamefont{and}
  \bibinfo{author}{\bibfnamefont{K.}~\bibnamefont{Levin}},
  \bibinfo{journal}{Phys. Rep.} \textbf{\bibinfo{volume}{412}},
  \bibinfo{pages}{1} (\bibinfo{year}{2005}).

\bibitem[{\citenamefont{Stajic et~al.}(2004)\citenamefont{Stajic, Milstein,
  Chen, Chiofalo, Holland, and Levin}}]{JS2}
\bibinfo{author}{\bibfnamefont{J.}~\bibnamefont{Stajic}},
  \bibinfo{author}{\bibfnamefont{J.~N.} \bibnamefont{Milstein}},
  \bibinfo{author}{\bibfnamefont{Q.~J.} \bibnamefont{Chen}},
  \bibinfo{author}{\bibfnamefont{M.~L.} \bibnamefont{Chiofalo}},
  \bibinfo{author}{\bibfnamefont{M.~J.} \bibnamefont{Holland}},
  \bibnamefont{and} \bibinfo{author}{\bibfnamefont{K.}~\bibnamefont{Levin}},
  \bibinfo{journal}{Phys. Rev. A} \textbf{\bibinfo{volume}{69}},
  \bibinfo{pages}{063610} (\bibinfo{year}{2004}).

\bibitem[{\citenamefont{Tan and Levin}(2004)}]{Tan2004}
\bibinfo{author}{\bibfnamefont{S.}~\bibnamefont{Tan}} \bibnamefont{and}
  \bibinfo{author}{\bibfnamefont{K.}~\bibnamefont{Levin}},
  \bibinfo{journal}{Phys. Rev. B} \textbf{\bibinfo{volume}{69}},
  \bibinfo{pages}{064510} (\bibinfo{year}{2004}).

\bibitem[{\citenamefont{Ullah and Dorsey}(1991)}]{Dorsey2}
\bibinfo{author}{\bibfnamefont{S.}~\bibnamefont{Ullah}} \bibnamefont{and}
  \bibinfo{author}{\bibfnamefont{A.}~\bibnamefont{Dorsey}},
  \bibinfo{journal}{Phys. Rev. B} \textbf{\bibinfo{volume}{44}},
  \bibinfo{pages}{262} (\bibinfo{year}{1991}).

\bibitem[{\citenamefont{Ussishkin et~al.}(2002)\citenamefont{Ussishkin, Sondhi,
  and Huse}}]{Huse}
\bibinfo{author}{\bibfnamefont{I.}~\bibnamefont{Ussishkin}},
  \bibinfo{author}{\bibfnamefont{S.}~\bibnamefont{Sondhi}}, \bibnamefont{and}
  \bibinfo{author}{\bibfnamefont{D.}~\bibnamefont{Huse}},
  \bibinfo{journal}{Phys. Rev. Lett.} \textbf{\bibinfo{volume}{89}},
  \bibinfo{pages}{287001} (\bibinfo{year}{2002}).

\bibitem[{\citenamefont{Maki}(1974)}]{Maki}
\bibinfo{author}{\bibfnamefont{K.}~\bibnamefont{Maki}}, \bibinfo{journal}{J.
  Low Temp. Physics} \textbf{\bibinfo{volume}{14}}, \bibinfo{pages}{419}
  (\bibinfo{year}{1974}).

\bibitem[{\citenamefont{Monroe et~al.}(1993)\citenamefont{Monroe, Cornell,
  Sackett, Myatt, and Wieman}}]{Monroe1993}
\bibinfo{author}{\bibfnamefont{C.~R.} \bibnamefont{Monroe}},
  \bibinfo{author}{\bibfnamefont{E.~A.} \bibnamefont{Cornell}},
  \bibinfo{author}{\bibfnamefont{C.~A.} \bibnamefont{Sackett}},
  \bibinfo{author}{\bibfnamefont{C.~J.} \bibnamefont{Myatt}}, \bibnamefont{and}
  \bibinfo{author}{\bibfnamefont{C.~E.} \bibnamefont{Wieman}},
  \bibinfo{journal}{Phys. Rev. Lett.} \textbf{\bibinfo{volume}{70}},
  \bibinfo{pages}{414} (\bibinfo{year}{1993}).

\bibitem[{\citenamefont{Guo et~al.}(2011{\natexlab{a}})\citenamefont{Guo,
  Wulin, Chien, and Levin}}]{Ourviscosity}
\bibinfo{author}{\bibfnamefont{H.}~\bibnamefont{Guo}},
  \bibinfo{author}{\bibfnamefont{D.}~\bibnamefont{Wulin}},
  \bibinfo{author}{\bibfnamefont{C.-C.} \bibnamefont{Chien}}, \bibnamefont{and}
  \bibinfo{author}{\bibfnamefont{K.}~\bibnamefont{Levin}},
  \bibinfo{journal}{Phys. Rev. Lett.} \textbf{\bibinfo{volume}{107}},
  \bibinfo{pages}{020403} (\bibinfo{year}{2011}{\natexlab{a}}).

\bibitem[{\citenamefont{Guo et~al.}(2011{\natexlab{b}})\citenamefont{Guo,
  Wulin, Chien, and Levin}}]{NJOP}
\bibinfo{author}{\bibfnamefont{H.}~\bibnamefont{Guo}},
  \bibinfo{author}{\bibfnamefont{D.}~\bibnamefont{Wulin}},
  \bibinfo{author}{\bibfnamefont{C.-C.} \bibnamefont{Chien}}, \bibnamefont{and}
  \bibinfo{author}{\bibfnamefont{K.}~\bibnamefont{Levin}},
  \bibinfo{journal}{New Journal of Physics} \textbf{\bibinfo{volume}{13}},
  \bibinfo{pages}{0i75011} (\bibinfo{year}{2011}{\natexlab{b}}).

\bibitem[{\citenamefont{Turlapov et~al.}(2008)\citenamefont{Turlapov, Kinast,
  Clancy, Luo, Joseph, and Thomas}}]{ThomasJLTP08}
\bibinfo{author}{\bibfnamefont{A.}~\bibnamefont{Turlapov}},
  \bibinfo{author}{\bibfnamefont{J.}~\bibnamefont{Kinast}},
  \bibinfo{author}{\bibfnamefont{B.}~\bibnamefont{Clancy}},
  \bibinfo{author}{\bibfnamefont{L.}~\bibnamefont{Luo}},
  \bibinfo{author}{\bibfnamefont{J.}~\bibnamefont{Joseph}}, \bibnamefont{and}
  \bibinfo{author}{\bibfnamefont{J.~E.} \bibnamefont{Thomas}},
  \bibinfo{journal}{J. Low Temp. Phys.} \textbf{\bibinfo{volume}{150}},
  \bibinfo{pages}{567} (\bibinfo{year}{2008}).

\bibitem[{\citenamefont{Kovtun et~al.}(2005)\citenamefont{Kovtun, Son, and
  Starinets}}]{Qviscosity}
\bibinfo{author}{\bibfnamefont{P.~K.} \bibnamefont{Kovtun}},
  \bibinfo{author}{\bibfnamefont{D.~T.} \bibnamefont{Son}}, \bibnamefont{and}
  \bibinfo{author}{\bibfnamefont{A.~O.} \bibnamefont{Starinets}},
  \bibinfo{journal}{Phys. Rev. Lett.} \textbf{\bibinfo{volume}{94}},
  \bibinfo{pages}{111601} (\bibinfo{year}{2005}).

\bibitem[{\citenamefont{C.~Cao and Thomas}(2011)}]{ThomasViscosity}
\bibinfo{author}{\bibfnamefont{J.~J. H. W. J. P. T.~S.} \bibnamefont{C.~Cao},
  \bibfnamefont{E.~Elliot}} \bibnamefont{and}
  \bibinfo{author}{\bibfnamefont{J.~E.} \bibnamefont{Thomas}},
  \bibinfo{journal}{Science} \textbf{\bibinfo{volume}{331}},
  \bibinfo{pages}{58} (\bibinfo{year}{2011}).

\bibitem[{\citenamefont{Zhang et~al.}(2012)\citenamefont{Zhang, Hung, Tung, and
  Chin}}]{Zhang2012}
\bibinfo{author}{\bibfnamefont{X.}~\bibnamefont{Zhang}},
  \bibinfo{author}{\bibfnamefont{C.-L.} \bibnamefont{Hung}},
  \bibinfo{author}{\bibfnamefont{S.-K.} \bibnamefont{Tung}}, \bibnamefont{and}
  \bibinfo{author}{\bibfnamefont{C.}~\bibnamefont{Chin}},
  \bibinfo{journal}{Science} \textbf{\bibinfo{volume}{335}},
  \bibinfo{pages}{1070} (\bibinfo{year}{2012}).

\end{thebibliography}

\end{document}